\documentclass{ws-ijmpa4hep}

\newcommand{\tr}{\mbox{tr}}

\def\lb{\label}

\def\fsl#1{\setbox0=\hbox{$#1$}                 
   \dimen0=\wd0                                 
   \setbox1=\hbox{/} \dimen1=\wd1               
   \ifdim\dimen0>\dimen1                        
      \rlap{\hbox to \dimen0{\hfil/\hfil}}      
      #1                                        
   \else                                        
      \rlap{\hbox to \dimen1{\hfil$#1$\hfil}}   
      /                                         
   \fi}                                         %

\begin{document}

\markboth{Michio Hashimoto}
{Noncommutativity vs. Transversality in QED in a strong magnetic field}

%
\catchline{}{}{}{}{}
%

\title{Noncommutativity vs. Transversality in QED in a strong magnetic field
\footnote{Based on talk given at {\it The 27th Annual
  Montreal-Rochester-Syracuse-Toronto Conference on High Energy
  Physics (MRST2005),} SUNY Institute of Technology, Utica, New York, 
  May 16--18, 2005.}}

\author{Michio Hashimoto}
\address{
Department of Applied Mathematics, University of Western
Ontario, \\ London, Ontario N6A 5B7, Canada\\
E-mail: mhashimo@uwo.ca}

\maketitle

\begin{history}
\end{history}


\begin{abstract}
Quantum electrodynamics (QED) in a strong constant magnetic field 
is investigated from the viewpoint of its connection with 
noncommutative QED.
It turns out that within the lowest Landau level (LLL) approximation
the 1-loop contribution of fermions provides an effective action
with the noncommutative $U(1)_{\rm NC}$ gauge symmetry.
As a result, the Ward-Takahashi identities connected with 
the initial $U(1)$ gauge symmetry are broken down 
in the LLL approximation.
On the other hand, it is shown that the sum over the infinite number 
of the higher Landau levels (HLL's) is relevant
despite the fact that each contribution of the HLL is suppressed.
Owing to this nondecoupling phenomenon the transversality
is restored in the whole effective action.
The kinematic region where the LLL contribution is dominant
is also discussed.

\end{abstract}

\ccode{11.10.Nx, 11.15.-q, 12.20.-m}

\section{Introduction}

Quantum electrodynamics (QED) in a constant magnetic field 
has been thoroughly investigated since the classical 
papers\cite{HEW,Sch}. 
Following Ref.\refcite{Gorbar:2005az}, 
we here analyze this ``old'' subject from 
a ``new'' viewpoint connected with noncommutative QED. 
(For reviews of noncommutative field theories (NCFT), 
see Ref.\refcite{DNS}. Phenomenological issues of noncommutative QED
have been studied, for example, in Ref.\refcite{Riad:2000vy}.) 

In this article, some sophisticated features of the dynamics in QED
in a strong magnetic field are revealed. It is shown that
in the approximation with the lowest Landau
level (LLL) dominance the initial $U(1)$ gauge symmetry
in the fermion determinant is transformed
into the noncommutative $U(1)_{\rm NC}$ gauge symmetry. In this regime,
the effective action is intimately connected with that in
noncommutative QED and the original $U(1)$ gauge Ward-Takahashi 
identities are broken. 
In fact, this
dynamics yields a modified noncommutative QED
in which the UV/IR mixing\cite{MRS} is absent, similarly to the case of 
the NJL model in a strong magnetic field\cite{GM}. 
However, it is not the end of the story.
We show that the contribution of an infinite number of 
the higher Landau levels (HLL's) plays very important role 
for restoration of the original $U(1)$ gauge symmetry. 
Although the contribution of each HLL is suppressed in an infrared region, 
their cumulative contribution is not (a nondecoupling phenomenon). 
The situation is dramatically changed when the contribution of 
the HLL's is incorporated:
The transversality is restored.
We also indicate the kinematic region where the
LLL approximation is reliable.

\section{The LLL Approximation and Noncommutativity}

Let us study a problem in QED in a strong magnetic field $B$.
We consider the case with a large number of fermion flavors $N$
in order to justify the 1-loop approximation of fermions
in the sense of $1/N$ expansion.
We also choose the current mass $m$ of fermions satisfying 
the condition $m_{dyn}\ll m \ll \sqrt{|eB|}$,
where $m_{dyn}$ is the dynamical mass of fermions generated
in the chiral symmetric QED in a magnetic field\cite{QED1}. 
\footnote {The dynamical mass is
$m_{dyn} \simeq \sqrt{|eB|} \exp\left(-N\right)$
for a large running coupling $\tilde{\alpha}_{b} \equiv
N\alpha_{b}$
related to the magnetic scale $\sqrt{|eB|}$, and
$m_{dyn} \sim \sqrt{|eB|}\exp\left[-\frac{\pi N}
{\tilde{\alpha}_{b}\ln\left(1/\tilde{\alpha}_{b}\right)}\right]$
when the coupling $\tilde{\alpha}_{b}$ is weak\cite{QED1}.}
The condition $m_{dyn}\ll m$ guarantees that there are no light
(pseudo) Nambu-Goldstone bosons, and the only particles in low energy 
are photons in this model. 
As to the condition  $m \ll 
\sqrt{|eB|}$,
it implies that the magnetic field is very strong.

Integrating out fermions, we obtain
the effective action for photons in the leading order in $1/N$:
\begin{equation}
 \Gamma = \Gamma^{(0)} + \Gamma^{(1)},
 \lb{ACTION}
\end{equation}
with the tree level part,
\begin{equation}
 \Gamma^{(0)} = - \frac{1}{4}\int d^4x \,f_{\mu\nu}^2,
 \lb{ACTION-tree}
\end{equation}
and the 1-loop part,
\begin{equation}
 \Gamma^{(1)} =
 -iN\mbox{Tr}\mbox{Ln} \left[ i\gamma^{\mu} (\partial_{\mu} 
 -ieA_{\mu}) -m\, \right],
 \lb{ACTION-1-loop}
\end{equation}
where $f^{\mu\nu} = \partial_{\mu}A_{\nu} - \partial_{\nu}A_{\mu}$
and the vector field $A_{\mu} = A^{cl}_{\mu} + \tilde{A}_{\mu}$.
The classical part $A^{cl}_{\mu}$ is 
$A^{cl}_{\mu} = \langle 0|A_{\mu}|0 \rangle$. 
For a constant magnetic field directed in the $+x^3$
direction, we may use the so called symmetric gauge for $A_{\mu}^{cl}$,
\begin{equation}
A_{\mu}^{cl} = (0,\frac{Bx^2}{2},-\frac{Bx^1}{2},0).
\label{symm}
\end{equation}

In a constant magnetic field the fermion propagator is given by\cite{Sch}
\begin{eqnarray}
S(x,y)=\exp\biggl[\frac{ie}{2}(x-y)^\mu A_\mu^{cl}(x+y)\biggr]
\tilde S(x-y)~, 
\label{propagator}
\end{eqnarray}
where the Fourier transform of the translationally invariant part 
$\tilde S$ can be decomposed over the Landau levels\cite{Chod}:
\begin{equation}
\tilde S(k)=i \exp\biggl(-\frac{{\bm k}_{\perp}^2}{|eB|}\biggr)
\sum^\infty_{n=0}
(-1)^n \frac{D_n(eB,k)}{k_{\|}^{2}-m^2-2n|eB|}
\lb{decomposition}  
\end{equation}
with ${\bm k}_{\perp}\equiv (k^1,k^2)$
and $k_{\|} \equiv (k_0, k_3)$. 
The functions $D_n(eB,k)$ are expressed through the
generalized Laguerre polynomials $L_m^\alpha$:
\begin{align}
D_n(eB,k) & \,=\, (k_{\|}\gamma^{\|} +m)
\biggl[\bigl(1-i\gamma^1\gamma^2 {\rm sign}(eB)\bigr)L_n
\biggl(2\frac{{\bm k}_{\perp}^2}{|eB|}\biggr) \nonumber \\
& \hspace*{2.5cm}
-\bigl(1+i\gamma^1\gamma^2{\rm sign}(eB)\bigr)L_{n-1}
\biggl(2\frac{{\bm k}_{\perp}^2}{|eB|}\biggr)\biggr]
 \nonumber\\ & \quad 
+4(k^1 \gamma^1+k^2\gamma^2)L^1_{n-1}
\biggl(2\frac{{\bm k}_{\perp}^2}{|eB|}\biggr)\, ,
\lb{L}
\end{align}
where $\gamma^{\|} \equiv (\gamma^0,\gamma^3)$.
For a strong magnetic field $|eB| \gg m^2$, 
we expect that in the infrared region
the LLL approximation should be reliable. 
Actually, the relation (\ref{decomposition}) seems to 
suggest that in the infrared region, $k_{\perp}, k_{\|} \ll \sqrt{|eB|}$, 
all the HLL's with $n \geq 1$ decouple because of their heavy mass 
$\sqrt{2n|eB|}$ and only the LLL with $n = 0$ is relevant.
Although the above argument is physically convincing, there may be a
potential flaw due to an infinite number of the Landau levels. 
As will be shown, it is indeed the case in this problem:
it turns out that the cumulative contribution of the HLL's does not decouple. 

Nevertheless, we first study the QED dynamics in the LLL approximation 
where the fermion propagator is replaced by
the LLL one in the calculation of the effective action (\ref{ACTION-1-loop}). 
Recently, the NJL model in a strong magnetic field has been
analyzed within the LLL approximation\cite{GM}. 
The extension of the analysis to the case of QED is straightforward. 
The effective action (\ref{ACTION}) in the LLL approximation is given by
\begin{equation}
 \Gamma_{\rm LLL} = \Gamma^{(0)} + \Gamma^{(1)}_{\rm LLL},
 \lb{starACTION}
\end{equation}
with
\begin{equation}
 \Gamma^{(1)}_{\rm LLL} = -\frac{iN|eB|}{2\pi} \int d^2x_{\perp}\,
 \mbox{Tr}_{||}\Big[{\cal P}\,
 \mbox{Ln}[i\gamma^{||}(\partial_{||} - ie{\cal A}_{||})-m]\Big]_{*} \;,
 \lb{starACTION-1-loop}
\end{equation}
(compare with Eq. (54) in Ref. \refcite{GM}). Here $*$ is the 
symbol of 
the Moyal star product, which is a signature of a NCFT\cite{DNS}, 
the spin projector ${\cal P}$ is defined by
\begin{equation}
 {\cal P} \equiv
  \frac{1}{2}\bigg[\,1 - i\gamma^1\gamma^2 \mbox{sign}(eB)\,\bigg],
\end{equation}
and the longitudinal ``smeared'' fields 
${\cal A}_{||}$ are defined as\cite{GM}  
\begin{equation}
 {\cal A}_{||} = e^{\frac{\nabla_{\perp}^2}{4|eB|}} A_{||}, 
\end{equation}
where $\nabla_{\perp}^2$ is the transverse Laplacian.
Notice that ${\cal P}$ is the
projector on the fermion (antifermion) states with the spin
polarized along (opposite to) the magnetic field
and that the one-loop term $\Gamma^{(1)}_{\rm LLL}$ in
(\ref{starACTION}) includes only the longitudinal field
${\cal A}_{||}  =
({\cal A}_{0}, {\cal A}_{3})$. This is because
the LLL fermions couple only to the longitudinal components of
the photon field\cite{QED1}.

In the effective action (\ref{starACTION-1-loop}), 
the trace $\mbox{Tr}_{||}$ of the longitudinal subspace should be
taken in the functional sense and the star product 
relates to the space transverse coordinates. Therefore
the LLL dynamics 
determines a NCFT with noncommutative
transverse coordinates $\hat{x}^{a}_{\perp}$, $a = 1,2$:
\begin{equation}
[\hat{x}^{a}_{\perp}, \hat{x}^{b}_{\perp}] = 
i\frac{1}{eB}\epsilon^{ab} \equiv i\theta^{ab}.
\label{commrel}
\end{equation}

The structure of the logarithm of the fermion determinant
in $\Gamma^{(1)}_{\rm LLL}$ implies that it is invariant not under
the initial $U(1)$ gauge symmetry but under the noncommutative
$U(1)_{\rm NC}$ gauge one\cite{DNS} 
(henceforth we omit the subscript $||$ in gauge fields):   
\begin{subequations}
\begin{align}
{\cal A}_{\mu} &\;\to\; U(x)*{\cal A}_{\mu}*U^{-1}(x) +
\frac{i}{e}U(x)*\partial_{\mu}U^{-1}(x), && (\mu=0,3)  \\[2mm]
{\cal F}_{\mu\nu} &\;\to\; U(x)*{\cal F}_{\mu\nu}*U^{-1}(x),
&& (\mu,\nu=0,3) 
\label{gt}
\end{align}
\end{subequations}
where $U(x)=(e^{i\lambda(x)})_{*}$ and the field strength 
${\cal F}_{\mu\nu}$ is
\begin{equation}
{\cal F}_{\mu\nu} = 
\partial_{\mu}{\cal A}_{\nu} - \partial_{\nu}{\cal A}_{\mu}
-ie[{\cal A}_{\mu},{\cal A}_{\nu}]_{\rm MB}^{}
\label{F}  
\end{equation}
with the Moyal bracket 
\begin{equation}
[{\cal A}_{\mu},{\cal A}_{\nu}]_{\rm MB}^{} \equiv 
{\cal A}_{\mu}*{\cal A}_{\nu} - {\cal A}_{\nu}*{\cal A}_{\mu}.   
\end{equation}

Therefore the derivative expansion of 
$\Gamma^{(1)}_{\rm LLL}$ should be
expressed through terms with the star product of the field 
${\cal F}_{\mu\nu}$ 
and its covariant derivatives:
\begin{equation}
  \Gamma^{(1)}_{\rm LLL} = 
   a_0 S_{{\cal F}^2}^{} + a_1 S_{{\cal F}^3}^{}
 + a_2 S_{({\cal D}{\cal F})^2}^{}
 + a_3 S_{{\cal D}^2 {\cal F}^2}^{} + \cdots \,\,\,,
\lb{expansion}
\end{equation}
where
\begin{eqnarray}
  S_{{\cal F}^2}^{} &\equiv& 
 -\frac{1}{4}\int d^2 x_{\perp}^{} d^2 x_{\parallel}^{}\;
 {\cal F}_{\mu\nu}*{\cal F}^{\mu\nu}, \\
  S_{{\cal F}^3}^{} &\equiv& ie \int d^2 x_{\perp}^{} d^2 x_{\parallel}^{}\;
 {\cal F}_{\mu\nu} * {\cal F}^{\nu\lambda} * {\cal F}_\lambda^{\;\;\mu}, \\
  S_{({\cal D} {\cal F})^2}^{} &\equiv& \int d^2 x_{\perp}^{} 
  d^2 x_{\parallel}^{}\;
 {\cal D}_\lambda {\cal F}^{\lambda\mu} * 
 {\cal D}^\rho {\cal F}_{\rho\mu}, \\
  S_{{\cal D}^2 {\cal F}^2}^{} &\equiv& \int d^2 x_{\perp}^{} 
  d^2 x_{\parallel}^{}\;
 {\cal D}_\lambda {\cal F}_{\mu\nu} * 
 {\cal D}^\lambda {\cal F}^{\mu\nu},
\end{eqnarray}
and the covariant derivative of ${\cal F}_{\mu\nu}$ is
${\cal D}_\lambda {\cal F}_{\mu\nu} = 
  \partial_{\lambda}{\cal F}_{\mu\nu} 
-ie[{\cal A}_{\lambda},{\cal F}_{\mu\nu}]_{\rm MB}^{}$.
These are all independent operators which have the mass dimension four
and six. In particular, by using the Jacobi identity,
\begin{equation}
 [{\cal D}_\mu,[{\cal D}_\nu,{\cal D}_\lambda]_{\rm MB}^{}]_{\rm MB}^{} + 
 [{\cal D}_\nu,[{\cal D}_\lambda,{\cal D}_\mu]_{\rm MB}^{}]_{\rm MB}^{} +
 [{\cal D}_\lambda,[{\cal D}_\mu,{\cal D}_\nu]_{\rm MB}^{}]_{\rm MB}^{} = 0,
\end{equation}
and the relation
${\cal F}_{\mu\nu}=ie^{-1}[{\cal D}_\mu,{\cal D}_\nu]_{\rm MB}^{}$, 
one can easily check that the operator
$\int d^2 x_{\perp}^{}d^2 x_{\parallel}^{}\,
 {\cal D}_\lambda {\cal F}_{\mu\nu} * {\cal D}^\mu {\cal F}^{\nu\lambda}$ 
is not independent: it is equal to $-1/2\,S_{{\cal D}^2 {\cal F}^2}^{}$.

The coefficients $a_i$, $(i=0,1,2,3,\cdots)$ 
in Eq. (\ref{expansion}) can be found from the
$n$-point photon vertices 
\begin{eqnarray}
T_{\rm LLL}^{(n)} &=& i\frac{(ie)^{n} N|eB|}{2\pi n}
 \int d^2x^{\perp}\,d^2x^{||}_1 \cdots d^2x^{||}_n\, \nonumber \\ &&
\qquad \quad 
 \mbox{tr}\left[\,S_{||}(x^{||}_1-x^{||}_2)
 \fsl{{\cal A}}_{||}(x^{\perp},x^{||}_2)\,...\,
 S_{||}(x^{||}_n-x^{||}_1)
 \fsl{{\cal A}}_{||}(x^{\perp},x^{||}_1)\,\right]_{*}
 \label{nvertex}
\end{eqnarray}
by expanding the vertices in powers of external momenta,
where
\begin{equation}
  S_{\|}(x_{\|}) = \int \frac{d^2k_{\|}}{(2\pi)^2}
  e^{-ik_{\|}x^{\|}} \frac{i}{k_{\|}\gamma^{\|} - m}\,{ \cal P}
\end{equation}
and $\fsl{{\cal A}}_{||} \equiv \gamma^{||}{\cal A}_{||}$.
In particular, from the vertices $T_{\rm LLL}^{(2)}$ and $T_{\rm LLL}^{(3)}$, 
we find the coefficients $a_0$, $a_1$, $a_2$, and $a_3$ connected with
the operators of the dimension four and six in the 
derivative expansion (\ref{expansion}) of $\Gamma^{(1)}_{\rm LLL}$: 
\begin{equation}
 a_0 = \frac{\tilde{\alpha}}{3\pi}\frac{|eB|}{m^2}, \quad
 a_1 = \frac{1}{60m^2}a_0, \quad a_2 = -\frac{1}{10m^2}a_0, \quad a_3=0,
\lb{coeff}
\end{equation}
where $\tilde{\alpha} \equiv N\alpha = Ne^2/(4\pi)$
(since in the presence of
a magnetic field the charge conjugation symmetry is broken\footnote{
Noncommutative QED is also not $C$ invariant.\cite{Sheikh-Jabbari:2000vi}},
Furry's theorem does not hold and thereby the 3-point vertex appears).

Notice that the action $\Gamma_{\rm LLL}$ (\ref{starACTION}) 
determines a conventional noncommutative QED only in the case of an induced 
photon field, when the Maxwell term $\Gamma^{(0)}$ is absent.
When this term is present, the action also determines a NCFT,
however, this NCFT is different from the conventional ones considered in 
the literature. In particular, expressing the photon field $A_{\mu}$
through the smeared field ${\cal A}_{\mu}$ as
$A_{\mu} = e^{\frac{-\nabla_{\perp}^2}{4|eB|}} {\cal A_{\mu}}$, we
find that the propagator of the smeared field rapidly, as 
$e^{\frac{-p^{2}_{\perp}}{2|eB|}}$, 
decreases for large transverse momenta.
The form-factor $e^{\frac{-p^{2}_{\perp}}{2|eB|}}$
built in the smeared field reflects an inner structure of
photons in a magnetic field.
This feature leads to removing the UV/IR mixing in this NCFT
(compare with the analysis of the UV/IR mixing in Sec. 4 of Ref. 
\refcite{GM}). 

\section{Nondecoupling Effect of The HLL and Transversality}

The $U(1)$ gauge Ward-Takahashi identities imply that the $n$-point 
photon vertex
$T^{\mu_{1}...\mu_{n}}(x_{1},...,x_{n})$ should be transverse, i.e.,
$\partial_{\mu_j} T^{\mu_{1}...\mu_{n}}(x_{1},...,x_{n}) = 0$ 
($j=1,2,\cdots,n$).
It is easy to show that the 2-point vertex 
$T_{\rm LLL}^{\mu_1\mu_2}$
yielding the polarization operator is transverse indeed. 
Now let us turn to the 3-point vertex and show that it is not
transverse, i.e., the Ward-Takahashi identities connected with the initial
gauge $U(1)$ are broken in the LLL approximation. 

In the LLL approximation, the 3-point vertex in the momentum space
is given by
\begin{equation}
  T_{\rm LLL}^{\mu_1\mu_2\mu_3}(k_1,k_2,k_3) = Ne^3 \frac{|eB|}{2\pi}
  \sin\left(\frac{1}{2}\theta_{ab}k_{1\,\perp}^a k_{2\,\perp}^b\right)
  \Delta_{\rm LLL}^{\mu_{1\parallel}\mu_{2\parallel}\mu_{3\parallel}}
  (k_{1\,\parallel},k_{2\,\parallel},k_{3\,\parallel})
\lb{3-point}
\end{equation}
with\footnote{
The analytic expression of 
$\Delta_{\rm LLL}^{\mu_{1\parallel}\mu_{2\parallel}\mu_{3\parallel}}$
is given in terms of
the two dimensional version of the Passarino-Veltman
functions\cite{Passarino}. }
\begin{eqnarray}
\lefteqn{\hspace*{-5mm}
 \Delta_{\rm LLL}^{\mu_{1\parallel}\mu_{2\parallel}\mu_{3\parallel}}
 (k_{1\,\parallel},k_{2\,\parallel},k_{3\,\parallel})
 \equiv
} \nonumber \\[3mm] &&
 \int \frac{d^2 \ell_\parallel}{i(2\pi)^2}
 \frac{\tr\Big[\gamma^{\mu_1}_\parallel\,[(\fsl{\ell}-\fsl{k}_1)_\parallel+m]\,
               \gamma^{\mu_2}_\parallel\,[(\fsl{\ell}+\fsl{k}_3)_\parallel+m]\, 
               \gamma^{\mu_3}_\parallel\,(\fsl{\ell}_\parallel+m)\Big]}
      {(\ell_\parallel^2-m^2)[(\ell-k_1)_\parallel^2-m^2]
       [(\ell+k_3)_\parallel^2-m^2]}.
\end{eqnarray}
The argument of the sine in Eq. (\ref{3-point}) is the Moyal cross product 
with 
$\theta_{ab}
= \epsilon^{ab}/eB$ (see Eq. (\ref{commrel})).
It is easy to find that the divergence of the vertex (\ref{3-point})
is not zero,
\begin{equation}
k_{1\mu_1} T_{\rm LLL}^{\mu_1\mu_2\mu_3}(k_1,k_2,k_3) =
-\frac{2e}{i}\,
\sin\left(\frac{1}{2}\theta_{ab}k_{1\,\perp}^a k_{2\,\perp}^b\right)
\Big[\,\Pi_{\,\parallel}^{\mu_2\mu_3}(k_{2\,\parallel}) - 
       \Pi_{\,\parallel}^{\mu_2\mu_3}(k_{3\,\parallel})\,\Big],
\label{divergence}
\end{equation}
with
\begin{equation}
\Pi_{\,\parallel}^{\mu\nu}(k_{\parallel}) = i\frac{2\tilde{\alpha}|eB|}{\pi}
\left(\,g_{\,\parallel}^{\mu\nu} - 
\dfrac{k_{\parallel}^{\mu}k_{\parallel}^{\nu}}
{k_{\parallel}^2}\right) \Pi(k_{\parallel}^2),
\label{polarization}
\end{equation}
and
\begin{equation}
\Pi(k_{\parallel}^2) \equiv 
1 + \dfrac{2m^2}{k_{\parallel} ^2\sqrt{1-\frac{4m^2}
{k_{\parallel}^2}}}
        \ln\frac{\phantom{-}1 + \sqrt{1-\dfrac{4m^2}
{k_{\parallel}^2}}}
                {-1+\sqrt{1-\dfrac{4m^2}
{k_{\parallel}^2}}} \; ,
\end{equation}
where $\Pi_{\,\parallel}^{\mu\nu}$ is the polarization tensor
(apparently it is transverse).
We here defined $g_{\parallel}^{}=\mbox{diag}(1,-1)$.
Hence the original $U(1)$ gauge Ward-Takahashi identities are broken
in the LLL approximation. 

The origin of the violation of the transversality is obviously
the change of the symmetry:
the $T^{(n)}_{\rm LLL}$ vertices come from the 1-loop part
$\Gamma^{(1)}_{\rm LLL}$ of the effective action which is invariant 
under the noncommutative $U(1)_{\rm NC}$ gauge symmetry
\footnote{The effective action $\Gamma_{\rm LLL}$ (\ref{starACTION})
including both of the tree and 1-loop parts enjoys the longitudinal
$U(1)_{||}$ gauge symmetry with gauge parameters $\alpha(x^{||})$.
This $U(1)_{||}$ is a subgroup of the initial $U(1)$ and 
noncommutative $U(1)_{\rm NC}$.}.
Therefore the Ward-Takahashi identities for the vertices $T^{(n)}_{\rm LLL}$ 
should not reflect the initial $U(1)$ gauge symmetry,
but $U(1)_{\rm NC}$. 

However, it is clear that the full QED dynamics yields
the transverse vertices.
There should exist an additional contribution that restores 
the transversality broken in the LLL approximation.
Surprisingly, we will find that heavy (naively decoupled) HLL's
play very important role for the restoration.

\begin{figure}[tbp]
 \begin{center}
  \psfig{file=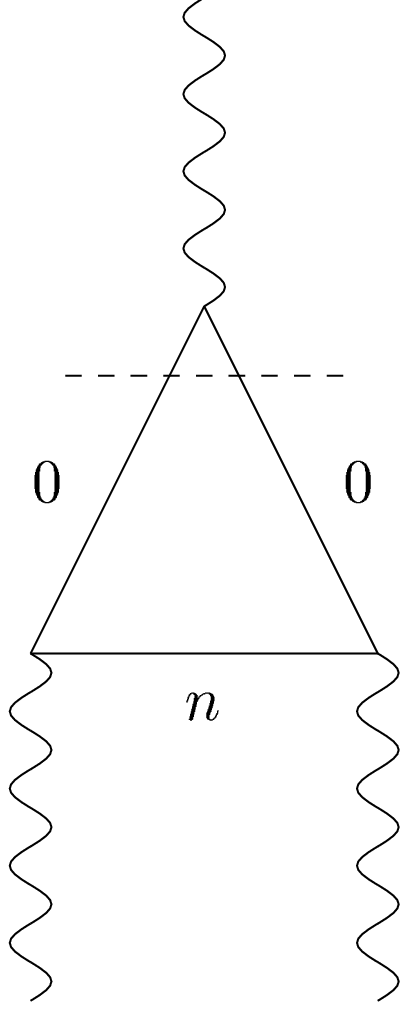,height=4.0cm}
 \end{center}
\vspace*{8pt}
\caption{A contribution of the HLL's. 
         Here ``$0$'' and ``$n$'' denote the fermion propagators of 
         the LLL and HLL with $n \geq 1$, respectively.
         This diagram clearly has a branch cut above $k_{||}^2 > 4m^2$.}
\label{fig1}
\end{figure}

\begin{figure}[tbp]
 \begin{center}
  \psfig{file=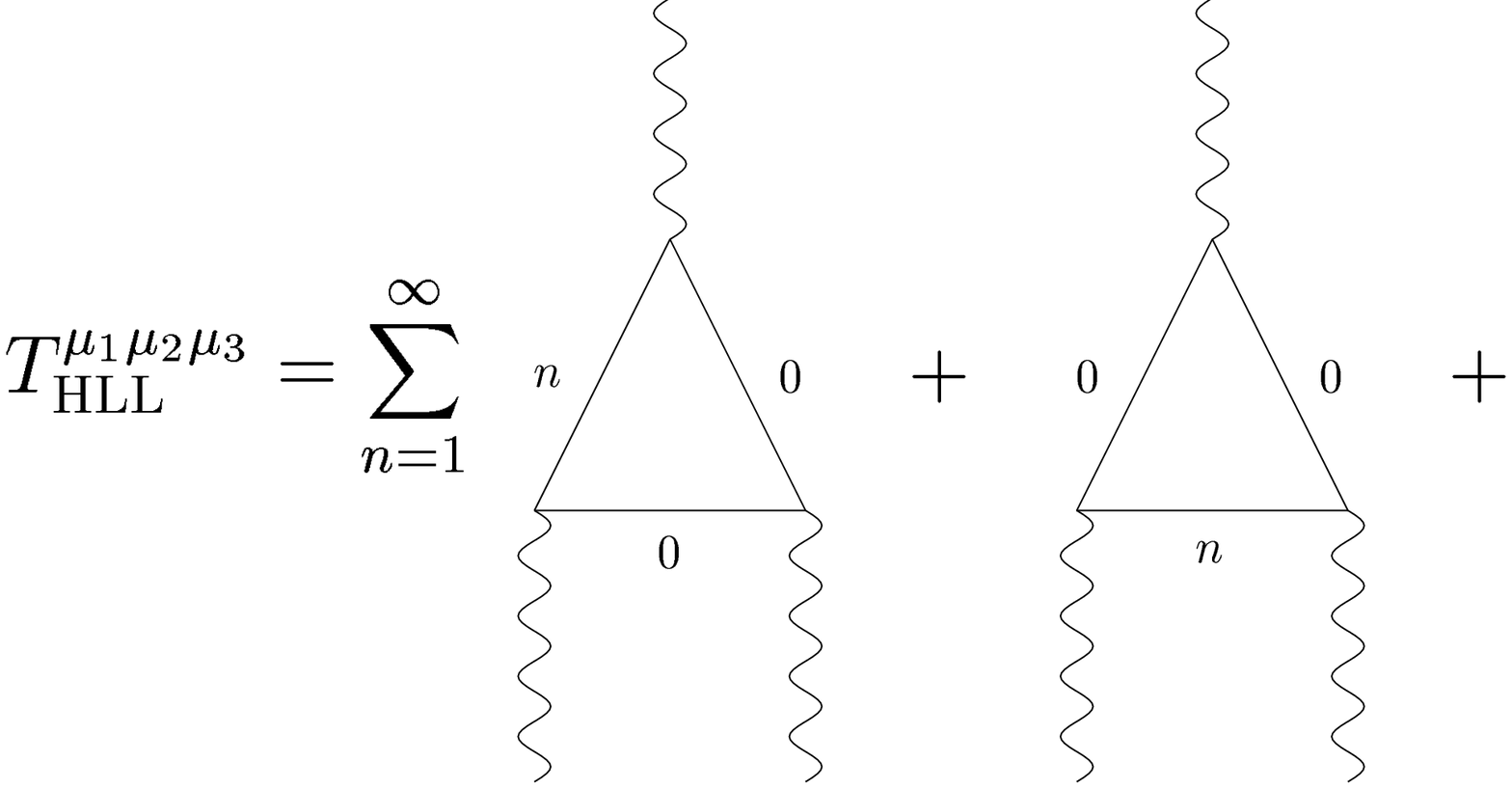,height=4.0cm}
 \end{center}
\vspace*{8pt}
\caption{The relevant contribution of the HLL's.}
\label{fig2}
\end{figure}

The 3-point vertex with the full fermion propagator (\ref{decomposition})
includes various contributions of the HLL's.
What kind of diagram is essential? 
We here note that the LLL diagram has a branch cut singularity 
above the threshold $k_{||}^2 > 4m^2$.
The relevant HLL contribution which we seek for should have 
the same analytic structure.
We easily find that only the diagram shown in Fig.\ref{fig1} 
has such a branch cut.
Therefore we consider the cumulative contribution of the particular
HLL diagrams shown in Fig.\ref{fig2} where one of the LLL propagator is 
replaced by the full one (\ref{decomposition}) without $n=0$.
By using a common technique in Ref.\refcite{tHooft:1978xw},
we can perform the loop integral with respect to $\ell_{||}$ and
hence schematically obtain\footnote{From a technical viewpoint, 
it is convenient to sum over the Laguerre polynomials 
before performing the loop integral with respect to $\ell_{\perp}$.
There is no subtlety concerning exchange of the ordering of
the infinite sum and the loop integral, because the integral 
is essentially a 2-dimensional one and does not have any UV divergence.}
\begin{equation}
  \Delta_{\rm HLL}^n \sim
  \frac{(-1)^n L^\alpha_n\biggl(2\frac{{\bm \ell}_{\perp}^2}{|eB|}\biggr)}
       {n|eB|} .
\end{equation}
After the integral over $\ell_{\perp}$, we find that 
the contribution of each of the individual HLL with $n \geq 1$ is 
\begin{equation}
 \Delta_{\rm HLL}^{\mu\nu\lambda}(p,q,k) =
 \frac{(-1)^n}{n|eB|}F^{\mu\nu\lambda}(p,q,k),
\end{equation}
where $F^{\mu\nu\lambda}$ is some function of longitudinal and transverse
momenta.
As was expected, each HLL contribution $\Delta_{\rm HLL}^{\mu\nu\lambda}$ 
is suppressed by powers of $1/|eB|$ in the infrared region.
It is, however, quite remarkable that despite the suppression of individual
HLL contributions, their cumulative contribution
becomes relevant in the infrared region.
In fact, by using the relation\cite{GR}
\begin{equation}
(1-z)^{-(\alpha+1)}\exp\biggl(\frac{xz}{z-1}\biggr)=\sum^\infty_{n=0}
L^\alpha_n(x)z^n~
\end{equation}
and integrating it with respect to $z$, 
we can perform explicitly the summation over the HLL contributions,
$\sum_{n=1}^\infty (-1)^n L^\alpha_n(x)/n$, and thereby
obtain a transverse vertex:
\begin{equation}
T^{\mu_1\mu_2\mu_3} (k_1,k_2,k_3) = 
  T^{\mu_1\mu_2\mu_3}_{\rm LLL}(k_1,k_2,k_3)
+ T^{\mu_1\mu_2\mu_3}_{\rm HLL}(k_1,k_2,k_3),
\lb{proper}
\end{equation}
where
\begin{eqnarray}
\lb{a}
T^{\mu_1\mu_2\mu_3}_{\rm HLL}(k_1,k_2,k_3) &=& \frac{2e}{i}\,
  \sin\left(\frac{1}{2}\theta_{ab}k_{1\,\perp}^a k_{2\,\perp}^b\right)
\Bigg[\,\dfrac{-k_{1\perp}^{\mu_1}}{{\bm k}_{1\perp}^2}\,
        \left(\,\phantom{\frac{}{}}
                \Pi_{\,\parallel}^{\mu_2\mu_3}(k_{2\parallel})
              - \Pi_{\,\parallel}^{\mu_2\mu_3}(k_{3\parallel})\,\right) 
\nonumber \\[3mm] && \quad
      + \frac{k_{1\perp}^{\mu_1}k_{2\perp}^{\mu_2}
             -k_{1\perp}^{\mu_2}k_{2\perp}^{\mu_1}
            -({\bm k}_1 \cdot {\bm k}_2)_\perp^{} g_{\,\perp}^{\mu_1\mu_2}}
             {{\bm k}_{1\perp}^2\,{\bm k}_{2\perp}^2}\,\, 
        k_{2\parallel\,\nu}\Pi_{\,\parallel}^{\mu_3\nu}(k_{3\parallel})
\nonumber \\[3mm] && \quad  
      + \dfrac{-k_{2\perp}^{\mu_2}}{{\bm k}_{2\perp}^2}\,
        \left(\,\phantom{\frac{}{}}
                \Pi_{\,\parallel}^{\mu_3\mu_1}(k_{3\parallel})
              - \Pi_{\,\parallel}^{\mu_3\mu_1}(k_{1\parallel})\,\right) 
\nonumber \\[3mm] && \quad
      + \frac{k_{2\perp}^{\mu_2}k_{3\perp}^{\mu_3}
             -k_{2\perp}^{\mu_3}k_{3\perp}^{\mu_2}
            -({\bm k}_2 \cdot {\bm k}_3)_\perp^{} g_{\,\perp}^{\mu_2\mu_3}}
             {{\bm k}_{2\perp}^2\,{\bm k}_{3\perp}^2}\,\, 
        k_{3\parallel\,\nu}\Pi_{\,\parallel}^{\mu_1\nu}(k_{1\parallel})
\nonumber \\[3mm] && \quad  
      + \dfrac{-k_{3\perp}^{\mu_3}}{{\bm k}_{3\perp}^2}\,
        \left(\,\phantom{\frac{}{}}
                \Pi_{\,\parallel}^{\mu_1\mu_2}(k_{1\parallel})
              - \Pi_{\,\parallel}^{\mu_1\mu_2}(k_{2\parallel})\,\right) 
\nonumber \\[3mm] && \quad
      + \frac{k_{3\perp}^{\mu_3}k_{1\perp}^{\mu_1}
             -k_{3\perp}^{\mu_1}k_{1\perp}^{\mu_3}
            -({\bm k}_3 \cdot {\bm k}_1)_\perp^{} g_{\,\perp}^{\mu_3\mu_1}}
             {{\bm k}_{3\perp}^2\,{\bm k}_{1\perp}^2}\,\, 
        k_{1\parallel\,\nu}\Pi_{\,\parallel}^{\mu_2\nu}(k_{2\parallel})\,\Bigg]
\nonumber \\[3mm] && + \,\mbox{(transverse part).}
\end{eqnarray}
We here defined $g_\perp^{}=\mbox{diag}(-1,-1)$, and 
$({\bm p} \cdot {\bm q})_\perp^{}=p^1q^1+p^2q^2$. 
Note that the vertex for the initial non-smeared fields $A_\mu$ 
is given by 
$e^{-\frac{{\bm k}_{1\perp}^2+{\bm k}_{2\perp}^2+{\bm k}_{3\perp}^2}{4|eB|}}
T^{\mu_1\mu_2\mu_3}$. 
It is easy to check the transversality of
the 3-point vertex $T^{\mu_1\mu_2\mu_3}$.
(See Eq.~(\ref{divergence}) and also note that the transversality of 
the vacuum polarization tensor,
$k_{||\,\mu}\Pi_{\,\parallel}^{\mu\nu}(k_{\parallel})=0$, 
and the momentum conservation, $k_1+k_2+k_3=0$.)

One might doubt whether or not 
there exists a kinematic region in which the LLL contribution 
is dominant.
We find a positive answer, i.e.,
the region with momenta ${\bm k}_{i\perp}^2 \gg |k_{i\parallel}^2|$.
In this region,
the leading terms in the expansion of the LLL and HLL vertices
in powers of $k_{i\parallel}$ are:
\begin{eqnarray}
  T^{\mu_1\mu_2\mu_3}_{\rm LLL}(k_1,k_2,k_3) &=& 
  -\frac{2e\tilde{\alpha}}{3\pi}\frac{|eB|}{m^2} 
   \sin\left(\frac{1}{2}\theta_{ab}k_{1\,\perp}^a k_{2\,\perp}^b\right)
   \Bigg[\,(k_2-k_3)_\parallel^{\mu_1} g_\parallel^{\mu_2\mu_3}
   \nonumber \\ && \qquad \qquad
          +(k_3-k_1)_\parallel^{\mu_2} g_\parallel^{\mu_3\mu_1}
          +(k_1-k_2)_\parallel^{\mu_3} g_\parallel^{\mu_1\mu_2}\,\Bigg], 
\end{eqnarray}
\begin{eqnarray}
\lefteqn{
  T^{\mu_1\mu_2\mu_3}_{\rm HLL}(k_1,k_2,k_3) = 
} \nonumber \\[3mm] &&
  -\frac{2e\tilde{\alpha}}{3\pi}\frac{|eB|}{m^2} 
  \sin\left(\frac{1}{2}\theta_{ab}k_{1\,\perp}^a k_{2\,\perp}^b\right)
  \Bigg[\,\dfrac{-k_{1\perp}^{\mu_1}}{{\bm k}_{1\perp}^2}\,
          \left(\,(k_{2\parallel}^2-k_{3\parallel}^2)g_\parallel^{\mu_2\mu_3}
              - k_{2\parallel}^{\mu_2}k_{2\parallel}^{\mu_3} 
              + k_{3\parallel}^{\mu_2}k_{3\parallel}^{\mu_3}
        \,\right) \nonumber \\[2mm] 
&& \qquad
      + \frac{k_{1\perp}^{\mu_1}k_{2\perp}^{\mu_2}
             -k_{1\perp}^{\mu_2}k_{2\perp}^{\mu_1}
            -({\bm k}_1 \cdot {\bm k}_2)_\perp^{} g_{\,\perp}^{\mu_1\mu_2}}
             {{\bm k}_{1\perp}^2\,{\bm k}_{2\perp}^2}\,
        \left(\,k_{3\parallel}^2 k_{2\parallel}^{\mu_3}
       -(k_2 \cdot k_3)_\parallel k_{3\parallel}^{\mu_3}\,\right) \nonumber \\[2mm]
&& \qquad + \,\, 
\mbox{permutations of}\,\, (k_1,\mu_1),\, (k_2,\mu_2),\,\,\mbox{and}\,\,
(k_3,\mu_3) \,\Bigg] \nonumber \\[3mm]
&&
 +\frac{2e\tilde{\alpha}}{3\pi}\frac{eB}{m^2} 
\Bigg[\,\left\{\,\exp\left(\,-\frac{({\bm k}_1\cdot{\bm k}_2)_\perp^{}}
                                   {2|eB|}\,\right)
        -\cos\left(\frac{1}{2}\theta_{ab}k_{1\,\perp}^a k_{2\,\perp}^b\right)
        \,\right\} \nonumber \\ && \qquad \times
\left\{\,\frac{k_{1\perp}^{\mu_2}\epsilon_\perp^{ab}k_{1\,\perp}^a k_{2\,\perp}^b
              +({\bm k}_1 \cdot {\bm k}_2)_\perp^{} \epsilon_\perp^{\mu_2 b}
               k_{1\,\perp}^b}
              {{\bm k}_{1\perp}^2\,{\bm k}_{2\perp}^2}\,\, 
          \left(\,k_{3\parallel}^2 g_\parallel^{\mu_3\mu_1} - 
                   k_{3\parallel}^{\mu_3}k_{3\parallel}^{\mu_1}\,\right)
\right. \nonumber \\ && \qquad \quad
        +\frac{k_{2\perp}^{\mu_1}\epsilon_\perp^{ab}k_{2\,\perp}^a k_{1\,\perp}^b
              +({\bm k}_2 \cdot {\bm k}_1)_\perp^{}
              \epsilon_\perp^{\mu_1 b}k_{2\,\perp}^b} 
              {{\bm k}_{1\perp}^2\,{\bm k}_{2\perp}^2}\,\, 
          \left(\,k_{3\parallel}^2 g_\parallel^{\mu_2\mu_3} - 
                   k_{3\parallel}^{\mu_2}k_{3\parallel}^{\mu_3}\,\right)
\nonumber \\ && \qquad \quad \left.
        +\frac{g_{\,\perp}^{\mu_1\mu_2}\epsilon_\perp^{ab}k_{1\,\perp}^a k_{2\,\perp}^b
              +\epsilon_\perp^{\mu_1 \mu_2}({\bm k}_1 \cdot {\bm k}_2)_\perp^{}}
              {{\bm k}_{1\perp}^2\,{\bm k}_{2\perp}^2}\,\,\,
        \left(\,k_{3\parallel}^2 k_{2\parallel}^{\mu_3}
       -(k_2 \cdot k_3)_\parallel k_{3\parallel}^{\mu_3}\,\right)\,\right\}
\nonumber \\ && \qquad + \,\, 
\mbox{permutations of}\,\, (k_1,\mu_1),\, (k_2,\mu_2),\,\,\mbox{and}\,\,
(k_3,\mu_3) \,\Bigg].
\end{eqnarray}
It is clear from these expressions that in that region the LLL 
contribution 
dominates indeed.
This result is quite noticeable.
The point is that as was shown in
Ref.\refcite{QED1}, the region with momenta ${\bm k}_{i\perp}^2 \gg 
|k_{i\parallel}^2|$ yields the dominant contribution in the
Schwinger-Dyson equation for the dynamical fermion mass in QED in a 
strong magnetic field. Therefore the LLL approximation is
reliable in that problem.

We comment on the role of $T^{\mu_1\mu_2\mu_3}_{\rm HLL}$. 
Although the vertex $T^{\mu_1\mu_2\mu_3}_{\rm HLL}$ is subdominant
in the region ${\bm k}_{i\perp}^2 \gg |k_{i\parallel}^2|$, 
it is crucial for restoration of the transversality.
While the LLL vertex $T^{\mu_1\mu_2\mu_3}_{\rm LLL}$ is multiplied 
only by a small longitudinal momentum $k_{\parallel}$,
the HLL vertex $T^{\mu_1\mu_2\mu_3}_{\rm HLL}$ includes certain terms
multiplied by a large transverse momentum $k_{\perp}$. 
Owing to this nature, the divergence of the subdominant term can 
cancel out the nonvanishing divergence of the dominant one.

\section{Summary}

We found that the LLL approximation yields the effective action
enjoying the noncommutative $U(1)_{\rm NC}$ gauge symmetry.
Hence the initial $U(1)$ gauge Ward-Takahashi identities are broken 
in the LLL approximation.
We also showed that nondecoupling phenomenon of (heavy) HLL's 
is the key point of the problem: 
the infinite sum over the HLL's is relevant to restoration of 
the transversality.

What physics underlines it? We believe 
that this phenomenon reflects the important role of a boundary dynamics  
at spatial infinity in this problem. The point is that the HLL's are
not only heavy states but their transverse size grows without limit with 
their gap $\sqrt{m^2  + 2n|eB|}$ as $n \to \infty$. This happens because
the transverse dynamics in the Landau problem is an oscillator-like one.
It implies that the role of the boundary dynamics at the 
transverse spatial infinity (corresponding to $n \to \infty$) is crucial.
This is similar to the role of edge states
in the quantum Hall effect: the edge states
are created by the boundary dynamics and also restore the gauge
invariance\cite{Hall}. 
Both these phenomena reflect the importance of
a boundary dynamics in a strong magnetic field. 
It would be interesting to examine whether or not similar
nondecoupling phenomena take place in noncommutative theories arising
in string theories in magnetic backgrounds\cite{DNS}.

\section*{Acknowledgments}
We are grateful for support from the Natural 
Sciences and Engineering Research Council of Canada.

\end{document}